\documentclass{emulateapj}
\def\gsim{\;\rlap{\lower 2.5pt
\hbox{$\sim$}}\raise 1.5pt\hbox{$>$}\;}
\def\lsim{\;\rlap{\lower 2.5pt
\hbox{$\sim$}}\raise 1.5pt\hbox{$<$}\;}

\begin{document}
\title{The Temperature of Interstellar Clouds from Turbulent Heating}

\author{Liubin Pan and Paolo Padoan}
\affil{Department of Physics, University of California, San Diego, 
CASS/UCSD 0424, 9500 Gilman Drive, La Jolla, CA 92093-0424; lpan@ucsd.edu, ppadoan@ucsd.edu}

\begin{abstract}

To evaluate the effect of turbulent heating in the thermal balance of interstellar clouds, we develop an extension of the log-Poisson 
intermittency model to supersonic turbulence. The model depends on a parameter, $d$, interpreted as the dimension of the most 
dissipative structures. By comparing the model with the probability distribution of the turbulent dissipation rate in a simulation of 
supersonic and super-Alfv\'{e}nic turbulence, we find a best-fit value of $d=1.64$. We apply this intermittency model to the computation 
of the mass-weighted probability distribution of the gas temperature of molecular clouds, high-mass star-forming cores, and cold diffuse 
HI clouds. Our main results are: i) The mean gas temperature in molecular clouds can be explained as the effect of turbulent heating alone, 
while cosmic ray heating may dominate only in regions where the turbulent heating is low; ii) The mean gas temperature in high-mass 
star-forming cores with typical FWHM of $\sim$ 6 km s$^{-1}$ (corresponding to a 1D rms velocity of 2.5 km s$^{-1}$)   
may be completely controlled by turbulent heating, which predicts a mean value of approximately 36~K,  two to 
three times larger than the mean gas temperature in the absence of turbulent heating; iii) The intermittency of the turbulent heating 
can generate enough hot regions in cold diffuse HI clouds to explain the observed CH$^+$ abundance, if the rms velocity on a scale 
of 1~pc is at least 3~km~s$^{-1}$, in agreement with previous results based on incompressible turbulence. Because of its importance 
in the thermal balance of molecular clouds and high-mass star-forming cores, the process of turbulent heating may be central in setting 
the characteristic stellar mass and in regulating molecular chemical reactions.

\end{abstract}


\section{Introduction}

The dissipation of turbulent kinetic energy provides a potentially important heating source in a variety of 
Galactic astrophysical environments, such as the solar wind (e.g., Matthaeus et al. 1999),  
interstellar clouds (e.g., Falgarone and Puget 1995), and the warm ionized medium (e.g., Minter and Balser 1997). 
The effect of turbulent heating has also been studied in extragalactic environments, such as in the context of the broad-line 
regions of quasars (Bottorff \& Ferland 2002), and in intracluster cooling flows (Dennis \& Chandran 2005). 

Astrophysical turbulence is often highly supersonic and magnetized. Energy decay in supersonic turbulence 
was thought to be very rapid due to shocks (e.g., Goldreich and Kwan 1974), while the presence of strong 
magnetic fields was believed to suppress the compressible modes and thus reduce the dissipation rate (Arons \& Max 1975). 
Recent numerical simulations offered a better understanding of energy dissipation in supersonic MHD turbulence
(Stone et al. 1998, Mac Low et al. 1998, Padoan and Norlund 1999). As in incompressible hydrodynamic 
turbulence, where the kinetic energy decays in a turnover time of the largest eddies, 
the dissipation timescale in supersonic MHD turbulence is of order the flow crossing timescale (Stone et al. 1998) 
or, equivalently, the dynamical timescale at the driving scale (Mac Low 1999). 

In typical molecular clouds the dynamical timescale is rather short, of order a million year, suggesting the need of 
continuous energy injection to support the observed turbulence. This result also implies a large turbulent heating rate. An 
estimate of the average dissipation rate from this timescale shows that it can be several times larger than the cosmic-ray 
heating rate, and thus may be the primary heating source in molecular clouds. Stone et al. (1998) also argued that the 
average turbulent heating rate can be comparable to the photoelectric heating in HI clouds with large velocity 
dispersions (however, according to Wolfire et al. 2003, the overall turbulent heating in the neutral medium may 
not be sufficient to produce the observed CII luminosity). In this paper, we provide a general theoretical formulation 
of the problem of turbulent heating, and investigate its effect on various types of interstellar clouds. 

Turbulent dissipation is characterized by its strong spatial roughness (see Figures~\ref{f1} and \ref{f2}). Extreme dissipation 
events appear in the smallest structures occupying a tiny volume or mass fraction, while a significant 
fraction of the flow experiences essentially no dissipation. This implies a broad probability distribution 
of the dissipation rate, which must be taken into account for a consistent investigation of turbulent heating. 
The extended tail of this distribution at large dissipation rates, corresponding to highly dissipative 
structures, is responsible for the anomalous scaling of the high-order velocity structure functions,
referred to as intermittency in turbulence theory (Frisch 1995). 

Intermittency has been extensively studied in incompressible turbulence. The intermittent model by 
She and Leveque (1994), which considers a hierarchy of dissipation rates of different levels 
and relates them to the fractal dimension of the most intermittent dissipative structures, has been very 
successful in reproducing the measured scaling exponents of structure functions in incompressible turbulence. 
It has been shown that the model is equivalent to a log-Poisson distribution of the dissipation rate 
(Dubrulle 1994, She and Waymire 1995). The probability distribution of the dissipation rate in supersonic turbulence 
has not been studied yet. Although the She and Leveque model with a fractal dimension of 2, corresponding to 
shocks, agrees well with the structure functions computed from numerical simulations of highly compressible 
turbulence (Boldyrev et al. 2002, Padoan et al. 2004), it remains to be confirmed 
whether the distribution of the dissipation rate in supersonic turbulence is consistent with a log-Poisson process. 
This theoretical concern and the wide application of an intermittency model of turbulent heating to various 
astrophysical environments are the primary motivations of the present work.   

An important effect of the intermittency of the turbulent dissipation is the generation of small regions with very large 
heating rate, and thus very high temperature. Falgarone and Puget (1995) were probably the first to
recognize the importance of this effect in cold HI clouds. Observed molecules, such as 
CH$^+$, suggest the existence of hot regions in cold HI clouds, because their production 
requires temperatures much higher than the average. Adopting experimental results from incompressible turbulent 
flows, they found that strong local turbulent heating in HI clouds could produce a sufficient fraction of 
hot regions to explain the observed abundance of CH$^+$ molecules. In this work we address the validity of their
result in the case of an intermittency model for supersonic turbulence, more appropriate
for cold HI clouds.  
      
In \S 2, we study the intermittent energy dissipation in supersonic turbulence. We show that the log-Poisson 
intermittency model gives probability distributions for the dissipation rate in excellent agreement 
with those at the resolved scales of numerical simulations of supersonic turbulence. In \S 3, we discuss heating and 
cooling processes in the interstellar medium and give the energy balance equations. We apply the log-Poisson 
intermittency model to investigate turbulent heating in molecular clouds, high-mass star-forming cores, 
and cold HI clouds in \S 4. Discussions and conclusions are given in \S 5.

\section{Intermittency of Turbulent Heating}

The energy dissipation rate in turbulent flows is known to exhibit fluctuations in both space and time. 
The viscous dissipation rate per unit mass is given by
\begin{equation}
\epsilon ({\bf x},t)
= \nu ( (\nabla \times {\bf v}) ^2 +\frac{4}{3} (\nabla \cdot {\bf v} )^2 )      
\label{eq1}
\end{equation}
where $\nu$ is the kinematic viscosity, and we have omitted the bulk viscosity, which is negligible 
for an ideal gas (a good approximation for the interstellar medium). The divergence term in eq.~(\ref{eq1}) is 
necessary for compressible flows. Clearly, in the presence of a fluctuating velocity field, 
the dissipation rate is expected to be inhomogeneous. Kolmogorov's 1941 theory uses the average 
dissipation rate (as the energy transfer rate in the inertial range) and thus implicitly assumes spatial 
homogeneity in the dissipation rate. That theory is therefore not sufficient to study turbulent heating. 
Because a significant fraction of the kinetic energy is viscously 
dissipated in the finest structures, such as vortex tubes and shocks (in supersonic turbulence), 
which occupy only a small volume fraction (or mass fraction for compressible turbulence),  the dissipation 
rate is strongly intermittent. Its probability distribution has a tail much more extended than that of a 
Gaussian distribution, as the result of strong local dissipative events. A careful consideration of this 
intermittent distribution is essential for investigating turbulent heating.  In the following, we briefly review
the intermittency theory.

\subsection{Intermittency Theory}
  
Kolmogorov (1962) and Oboukhov (1962) developed the first intermittency model for incompressible turbulence. 
To account for the fluctuations in the dissipation rate, they defined the dissipation rate, $\epsilon_l$, at each scale $l$ 
and considered its probability distribution as a function of $l$, which, together with the refined self-similarity hypothesis 
(Kolmogorov 1962), was used to predict the scaling exponents of the velocity structure functions.       
Generalizing to a compressible flow, the dissipation rate per unit mass at a scale $l$ is,   
\begin{equation}          
\epsilon_l ({\bf x},t) = \frac{1}{{\rho}_l({\bf x},t)V(l) } \int\limits_{|{\bf x'}|<l} \rho( {\bf x}+{\bf x'},t ) 
\epsilon ({\bf x}+{\bf x'},t) d{\bf x'}  
\label{eq2}
\end{equation} 
where $V(l)=4 \pi l^3/3$ is the volume of a spherical region of size $l$ and  
\begin{equation} 
{\rho}_l ({\bf x},t) = \frac{1}{V(l)} \int \limits_{ |{\bf x'}|<l } \rho( {\bf x}+{\bf x'},t ) d{\bf x'} 
\label{eq3}
\end{equation} 
is the average density of that spherical region. From its definition, $\epsilon_l$ is a function of ${\bf x}$ and we aim to study 
its spatial fluctuations. Note that the average rate, $\epsilon_l$, over a region of size $l$ defined here 
is different from $\nu v_l^2/l^2$ used by Falgarone and Puget (1995), which neglects the velocity gradient fluctuations 
in the region, and thus is valid only for $l$ smaller than the Kolmogorov dissipation scale, $\eta$, below which the 
velocity field is smooth and the velocity gradient is approximately constant. 

We point out that the pdf needed to study turbulent heating is that at the dissipation scale, the smallest scale 
of fluctuations in the dissipation rate. For this purpose, an ideal intermittency model to use is one for the dissipation 
range. However, most intermittency models, including the log-Poisson model we will adopt below, are for the inertial scales,  
and we are not aware of theoretical models that give the pdf of the dissipation rate in the dissipation range. 
Here, we take the following approach to obtain the pdf at the dissipation scale. We first focus on the inertial range and 
test an intermittency theory for the dissipation rate pdf in the inertial scales by comparing its predicted pdf at 
each scale with that computed from numerical simulations. If the theoretical pdf and its scale dependence agree 
with those from the simulation data, we extrapolate the model to obtain the pdf at the dissipation scale. This assumes that 
the pdf at the dissipation scale can be approximated by that extrapolated from the inertial scales. Although there is no 
theoretical estimate for its accuracy, the assumption is justified by the expectation that the pdf in the dissipation 
range probably connects continuously and smoothly to that in the inertial range and is supported by our 
numerical results (see below).  


The probability distribution, $P(\epsilon_l)$, of the dissipation rate, $\epsilon_l$, at a scale $l$ can be calculated as
\begin{equation}          
P(\epsilon_l) = \frac{1} { {\bar \rho} V} \int \delta ( \epsilon_l-\epsilon_l({\bf x},t) ) \rho_l( {\bf x},t ) d{\bf x} 
\label{eq4}
\end{equation} 
where $V$ is the total volume of the system and ${\bar \rho}$ is the overall average density in the flow. We have used the density as 
a weighting factor ($\rho_l/{\bar \rho}$, which is unity for incompressible turbulence) to account for the density variations in supersonic 
turbulence. This density or mass weighting is supported by Kritsuk et al. (2007), who find that replacing the velocity field 
${\bf v}$ by $\rho^{1/3} {\bf v}$ gives a 3rd order structure function in supersonic turbulence in agreement with Kolmogorov's 4/5 law. 
This is equivalent to saying that the density-weighted average dissipation rate, $\langle \epsilon_l \rangle$, is the same over all the inertial 
scales. This is built in our definition, eq.~(\ref{eq4}), 
\begin{equation}
\langle \epsilon_l \rangle = \int \epsilon_l P(\epsilon_l) 
d\epsilon_l \propto \int \rho_l({\bf x}, t) \epsilon_l({\bf x}, t) d {\bf x}
\label{eq5}
\end{equation}
which is independent of $l$, as follows from eq.~(\ref{eq2}). We will therefore denote the average dissipation rate 
$\langle \epsilon_l \rangle$ as ${\bar \epsilon}$. Clearly, ${\bar \epsilon} \simeq U^3/L$, where $U$ is the rms velocity of the flow 
and $L$ is the integral scale.       
Another reason for choosing the density weighted distribution is that we want to estimate the mass-weighted average temperature 
resulting from the turbulent heating of interstellar clouds. 
       
Kolmogorov (1962) and Oboukhov (1962) proposed a log-normal distribution for $\epsilon_l$, for which Yaglom (1966) gave 
a justification with a self-similar eddy-fragmentation argument. The variance of the log-normal distribution is 
assumed to be proportional to the number of cascade steps, $\propto$ ln$(L/l)$, from the integral scale, $L$, to the scale of interest, 
$l$. Experiments by Ansemet et al. (1984) find that the scaling exponents of the structure functions agree with the log-normal 
model at low orders, but depart from it at orders $\gsim 10$. The log-normal model 
has also been shown to violate some theoretical requirements on the structure function exponents in incompressible turbulence 
(Frisch 1995). Numerical simulations find that the log-normal distribution of the dissipation rate is quite a good approximation 
in the inertial range and is valid up to about 4-5 $\sigma$, beyond which a departure from the simulations is clearly seen (e.g., Yueng et al. 2006). 
This departure explains the deviation of the model from the measured structure functions in the experiments. In this work we do not consider 
the log-normal model (which was proposed for incompressible turbulence), because it gives a poor fit to $P(\epsilon_l)$ computed in our
simulations of supersonic turbulence. 

A different intermittency model has recently been proposed by She and Leveque (1994). The scaling exponents of 
structure functions of order up to 10 predicted by this model for incompressible hydrodynamic turbulence  
agree with experimental data with an accuracy of 1\%. In their original paper, She and Leveque start from the hierarchy of 
dissipation rates at each scale. By invoking a ``hidden symmetry'' that relates the dissipation rate at different intensity levels to 
the strongest dissipative structures at a given scale $l$, they determine the whole hierarchy of dissipation rates 
from the fractal dimension of the most intermittent structures and from the scaling of 
the dissipation rate of these structures with $l$. The ``hidden symmetry'' was speculated to be an unknown symmetry 
of the Navier-Stokes equation and was immediately recognized as corresponding to a log-Poisson process in 
a multiplicative energy cascade model. 

The model has been successfully applied to incompressible MHD turbulence (Muller and Biskamp 2000), supersonic turbulence 
(Boldyrev et al. 2002) and super-Alv\'{e}nic turbulent flows (Padoan et al. 2004). These studies compute the scaling exponents of the 
structure functions from numerical simulations and compare them to the prediction of the log-Poisson model with different fractal 
dimensions for the most intermittent structures. They find that a 2D geometry of the most intermittent structures, 
corresponding to current sheets in MHD turbulence or shocks in highly supersonic turbulence, gives an excellent fit to the simulation 
results. Here, instead of considering the scaling exponents of the structure functions, we focus on the log-Poisson version of the 
She-Leveque model and compare the model directly with the probability distribution, $P(\epsilon_l)$, measured in simulations 
of supersonic MHD turbulence. Our purpose is to check whether this model agrees with numerical simulations. If so, we can 
first ``calibrate'' it by fitting resolved inertial scales in the simulations, and then extrapolate it to high Reynolds number flows.   
In the following, we give a brief description of the log-Poisson model. The interested reader is referred to Dubrulle (1994) and 
to She and Waymire (1995) for more details (see also Pan et al. 2008, which includes both the original presentation by She 
and Leveque and the log-Poisson version).

\subsection{The Log-Poisson Model}

The log-Poisson model is a multiplicative cascade model where the dissipation rates at two scales, $l_1$ and $l_2$, are related by a 
multiplicative factor, $W_{l_1l_2}$, 
\begin{equation}
\epsilon_{l_2}=W_{l_2l_1} \epsilon_{l_1}
\label{eq6}
\end{equation}   
Since the average dissipation rate at each scale is equal to the overall average dissipation rate ${\bar \epsilon}$ (see above), 
the average of the multiplicative factor, $\langle W_{l_2l_1} \rangle $, is required to be unity. According to the speculation by 
She and Waymire (1995),  two types of events determine $W_{l_1l_2}$. One is the amplification of the dissipation rate that tends to 
cause singular structures at small scales. The other, called the modulation-defects by She and Waymire, corresponds to the failure 
of structures at scale $l_1$ to turn into the most dissipative structures at scale $l_2$ in the cascade. 
The first type of events is assumed to give an amplification factor of $(l_1/l_2)^\gamma$, which approaches infinity 
as $l_2 \to 0$ for positive $\gamma$. The parameter $\gamma$ measures how intense the most intermittent structures 
at scale $l_2$ can be. Assuming that the kinetic energy per unit mass available for dissipation in the most intermittent 
structures is $\sim U^2$, where $U$ is the rms velocity at the integral scale, and that the dynamical timescale $t_l$ in these 
structures follows the regular Kolmogorov scaling, $t_l \propto l^{2/3}$, She and Leveque (1994) argued that the energy 
dissipation rate in these intense structures is $ \sim U^2/t_l \propto l^{-2/3}$, which suggests that $\gamma=2/3$. 
The modulation-defects are assumed to be a discrete process, where each defect reduces $W_{l_1l_2}$ by a factor of $\beta$ and the number
of defects, $n$, occurring in the cascade from $l_1$ to $l_2$ is assumed to obey a Poisson distribution. The physical meaning of the 
parameter $\beta$ will be discussed below. These arguments are represented by two equations, 
\begin{equation}
W_{l_2l_1}=(l_1/l_2)^\gamma \beta^n
\label{eq7}
\end{equation}  
and 
\begin{equation}    
P(n)=exp(-\lambda_{l_1l_2}) \frac{\lambda_{l_1l_2}^n}{n!}
\label{eq8}
\end{equation}
where $\lambda_{l_1l_2}$ is the average number of defects in the cascade from $l_2$ to $l_1$. Requiring $\langle W_{l_1l_2} \rangle =1$ 
gives $\lambda_{l_1l_2}= {\gamma ln(l_1/l_2)} /(1-\beta)$. As expected, $\lambda_{l_1l_2}$ is proportional to the number of steps in the 
cascade, $ln(l_1/l_2)$.

\begin{figure}
\epsscale{1}
\plotone{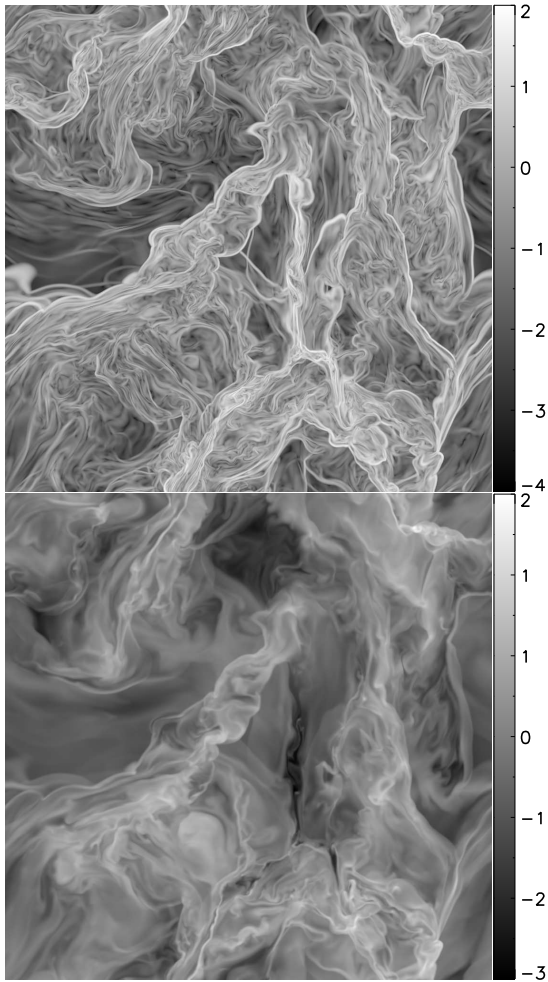} 
\caption{Logarithm of the dissipation rate (log($\epsilon({\bf x})/\bar{\epsilon}$), upper panel) and density 
(log($\rho({\bf x})/\bar{\rho}$), lower panel) relative to their averages on a slice of a snapshot of the simulation.
Both the dissipation rate and the density have a filamentary structure. Dense filaments are usually sites of strong dissipation,
but filaments of high dissipation rate are also found at the interface of low and high density regions. As a result, the dissipation 
rate and the gas density are practically uncorrelated (see Fig.~\ref{f3}).}
\label{f1}
\end{figure}

\begin{figure}
\epsscale{1}
\plotone{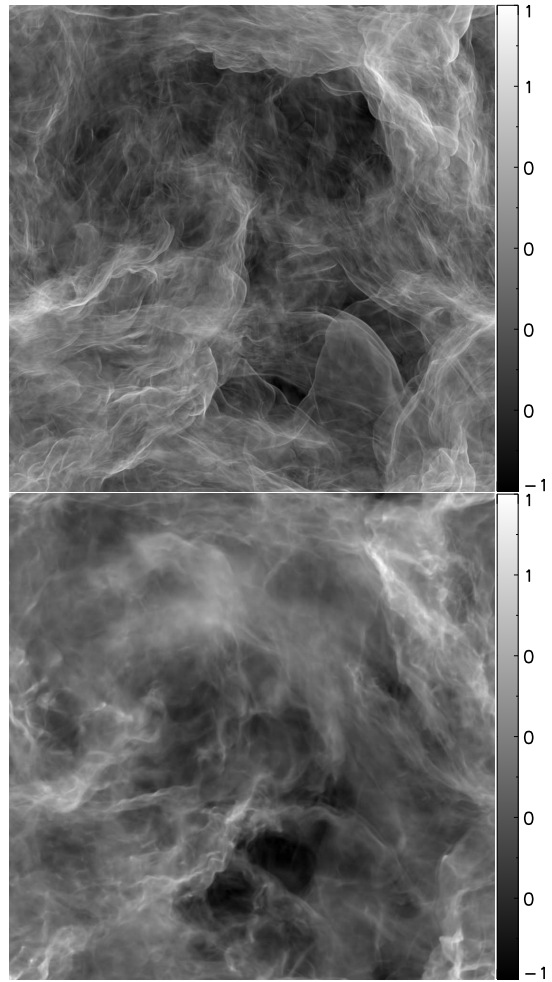} 
\caption{Logarithm of the projected dissipation rate (upper panel) and of the projected density (lower panel) relative to their averages, 
from a snapshot of the simulation.
As a result of the lack of correlation between the two quantities (see Fig.~\ref{f3}), individual structures in the projected dissipation rate do not 
have an obvious counterpart in the projected density.}
\label{f2}
\end{figure}

Taking the logarithm of eqs.~(\ref{eq6}) and (\ref{eq7}), we see that the distribution of $\epsilon_l$ is log-Poisson. Given the probability 
distribution at one scale, the distribution of $\epsilon_l$ at any other scale $l$ can be derived using eqs.~(\ref{eq6}), (\ref{eq7}) and (\ref{eq8}). 
The convenient scale to start with is the integral scale, $L$, where the distribution, $P_L$, of the dissipation rate 
mainly depends on the forcing of the flow and thus is probably flow-dependent. However, the dissipation rate, 
$\epsilon_L$, at $L$ is close to the average, $\bar \epsilon$,  and hence $P_L$ is expected to be narrow and approximately 
a delta function. Given $P_L (ln \epsilon_L)$, we find that 
 \begin{eqnarray}
\lefteqn{P(\epsilon_l) d\epsilon_l=}  \label{eq9} \\
& &  \sum\limits_{n=0}^{\infty} exp(-\lambda)\frac{\lambda^n}{n!}P_{L}(ln(\epsilon_{l}/\bar \epsilon)-\gamma ln(L/l)- n ln(\beta)) dln(\epsilon_l/\bar\epsilon) \nonumber 
\end{eqnarray}
where $\lambda=\lambda_{Ll}=\gamma ln(L/l)/(1-\beta)$. Note that if $P_L$ is a delta function, then $P(\epsilon_l)$ is discrete 
with a series of spikes. 

The physical meaning of the parameter $\beta$ in the log-Poisson version of the model is not clear. Deriving the moments of the distribution 
$P(\epsilon_l)$ from eq.~(\ref{eq9}) and comparing with the original version of the model, one finds that $\beta$ here corresponds 
to the same $\beta$ parameter introduced in the ``hidden symmetry'' by She and Leveque (1994), which has been related to the 
fractal dimension of the most intermittent structures, $d$, and the parameter $\gamma$ by $\gamma/(1-\beta) = D -d $, with $D=3$ 
being the dimension of the system. Here we omit the derivation of this relation that can be found, e.g., in Pan et al. (2008). 
For vortex tubes, the dimension is $d=1$ and for shocks $d=2$. Because in supersonic turbulence about 1/3 of the energy is 
dissipated in dilatational modes and 2/3 in solenoidal modes (Kritsuk et al. 2007), $d$ is expected to be between 1 and 2.

\subsection{The Log-Poisson Model for Supersonic Turbulence}

We compare the log-Poisson distribution, eq.~(\ref{eq9}), with results from a numerical simulation of supersonic MHD turbulence.  
We examine how well this model can be extended to supersonic MHD flows and what fractal dimension of the most intense 
dissipative structures gives the best fit to the simulation. We take 6 snapshots from the 1000$^3$ Stagger-code MHD simulation 
in Padoan et al. (2007). The simulation adopts periodic boundary conditions, isothermal equation of state, random forcing in Fourier 
space at wavenumbers $1\le k\le 2$ ($k=1$ corresponds to the computational box size), uniform initial density and magnetic field, 
and random initial velocity field with power only at wavenumbers $1 \le k\le 2$.  The simulation is both supersonic and super-Alfv\'{e}nic,
with an rms sonic Mach number of ${\cal M}_{\rm s} \approx 9$, and an initial rms Alfv\'{e}nic Mach number of ${\cal M}_{\rm a,i}=29.7$.
After a few dynamical times, the magnetic energy is amplified by the turbulence, and the rms Alfv\'{e}nic Mach number with respect to the 
rms magnetic field is ${\cal M}_{\rm a}\approx 2.8$, still super-Alfv\'{e}nic. 

In each snapshot, we evaluate the dissipation rate at the grid points according to eq.~(\ref{eq1}). 
A complication arises as a result of the limited numerical resolution. The dissipation rate should be calculated at the 
Kolmogorov scale using the kinematic viscosity, $\nu \simeq c/(n \sigma)$, where $c$ is the sound speed, $n$ the number density 
and $\sigma$ the collision cross section. This is not possible because the resolution scale in the simulation is much larger than the 
Kolmogorov scale. We cannot use the kinematic viscosity to calculate the dissipation rate at the grid points, 
because that neglects the fluctuations in velocity gradients below the grid scale and would underestimate the dissipation rate 
in each computational cell. To avoid this, we are forced to use an effective viscosity at the grid size to account for the sub-grid fluctuations 
erased by the numerical viscosity. In principle, the effective viscosity at each grid depends on the sub-grid fluctuations and 
(perhaps weakly) on the local Kolmogorov scale (which varies in space because the kinematic viscosity $\nu$ depends on density). 
Due to the lack of information about sub-grid scales, we simply assume the effective viscosity is constant, and normalize it so that the mean 
dissipation rate averaged over all the grid points is equal to the overall dissipation rate of the flow, $\bar \epsilon \simeq U^3/L$. 
With this assumption, we calculate the dissipation rate in each computational cell.

\begin{figure}
\epsscale{1.0}
\plotone{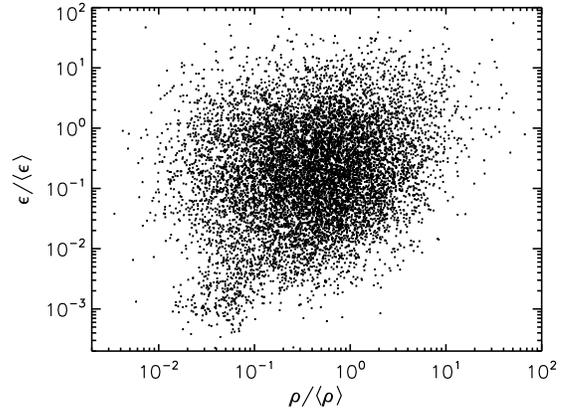} 
\caption{Dissipation rate versus gas density in 10,000 randomly sampled computational cells from one snapshot of the simulation.
The dissipation rate and the gas density are practically uncorrelated. The correlation coefficient computed over all the computational
cells of one snapshot is -0.11.}
\label{f3}
\end{figure}

Figure~\ref{f1} shows the logarithm of the dissipation 
rate (upper panel) and density (lower panel) on a slice of a snapshot of the simulation. Both the dissipation rate and the density have a 
filamentary structure. Dense filaments are usually sites of strong dissipation, but filaments of high dissipation rate are also found at the 
interface of low and high density regions. As a result, the dissipation rate and the gas density are practically uncorrelated. This lack of
statistical correlation between dissipation rate and gas density is apparent when comparing images of projected dissipation rate and
projected density, shown in Figure~\ref{f2}. Structures in these two images appear to be rather independent of each other. 
The correlation coefficient between dissipation rate and gas density, computed over all the computational cells of one snapshot, is very
low, -0.11, as illustrated by the scatter plot of Figure~\ref{f3}, for 10,000 randomly selected computational cells of one snapshot. 

From the dissipation rate at each computational cell, we compute the average dissipation rate, $\epsilon_l$, at any larger scale $l$, 
using eq.~(\ref{eq2}) (instead of calculating the average over spheres, we compute that over cubes of size $l$ from the simulation data). 
We then calculate the density-weighted probability distribution, $P(\epsilon_l)$, of the average dissipation rate, $\epsilon_l$, over 
regions of size $l$, using eq.~(\ref{eq4}). We carry out the same calculation for each snapshot and average the distributions 
from the 6 snapshots. The results are plotted in Figure~4 for $l=L/64$, $L/128$, $L/256$, and $L/512$, where $L$
is the size of the simulation box. The distribution becomes broader toward smaller scales, indicating the 
dissipation rate is more intermittent at smaller scales. Note that the distribution of $\log(\epsilon)$ is strongly skewed toward small 
values, suggesting that a log-normal model cannot give a satisfactory fit to the numerical results.

Figure~4 also shows the probability distribution from the log-Poisson model, eq.~(\ref{eq9}), 
with $\gamma=2/3$ and $\beta=0.51$ (which corresponds to a fractal dimension of $d=1.64$ for the most intense dissipative structures) 
at the same scales $l$. As mentioned above, at the box size $L$, $\epsilon_L=\bar \epsilon$ and $P_L$ is close to a delta function. 
We have set $P_L$ to be Gaussian with a small variance that is just large enough to make the distribution given by eq.~(\ref{eq9}) a 
smooth function. The log-Poisson model with a fractal dimension of the most dissipative structures $d=1.64$ gives an excellent agreement 
with numerical simulations over more than 5 orders of magnitudes in the dissipation rate. The agreement for $l=L/128$ and $L/256$ is 
especially remarkable, while slight departures for $l= L/64$ and $L/512$ are perhaps due to, respectively, the memory of large-scale 
motions and the suppression of intermittency by numerical dissipation. Note that the motivation of the log-Poisson model was 
to explain the tail at large dissipation rates corresponding to high-order structure functions. It turns out that this model provides a
surprisingly good fit to the tail of low dissipation rates as well. Padoan et al. (2004) obtained a fractal dimension of about 2 from 
MHD simulations with similar Mach numbers as here. In that paper, the structure functions were computed without density weighting, 
which may explain the difference in the resulting dimension of the dissipative structures. 
Kritsuk et al. (2007) found that two dimensions, $d=1.5$ and $d=2.25$, can give acceptable fit to the scaling exponents for 
the structure functions of $\rho^{1/3}{\bf v}$ in supersonic hydrodynamic turbulence.  
From their Fig. 2, the fitting quality is better for $d=1.5$, which is quite close to our result here. However, we point out that, 
except for the third order structure function (corresponding to the first order moment of the dissipation rate here), 
the density-weighting in $\rho^{1/3}{\bf v}$ is different from that in eq.~ (\ref{eq4}).

\begin{figure}
\epsscale{1.0}
\plotone{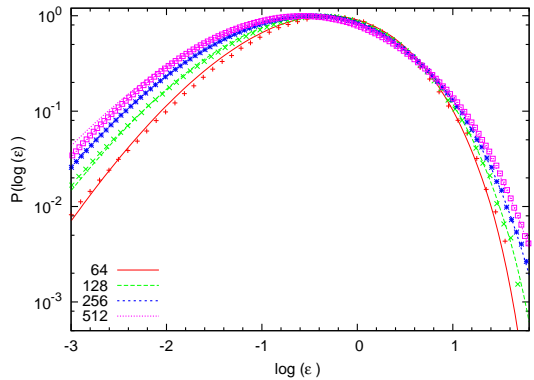}
\caption{Density-weighted probability distribution of the dissipation rate at different scales. 
The symbols are from our simulation of supersonic MHD turbulence, and the curves from the 
log-Poisson model (eq.~(\ref{eq9})) with $\gamma=2/3$ and $\beta=0.51$, corresponding to a fractal dimension 
$d=1.64$ of the most intermittent structures.  The model gives an excellent fit to the distribution computed from 
the simulation. From bottom to top, the different symbols and the curves correspond to the distribution at 
$L/64$, $L/128$, $L/256$, and $L/512$, where $L$ is the computational box size. The distribution becomes broader 
toward smaller scales, indicating stronger intermittency at smaller scales.             
\label{f4}}
\end{figure}

To evaluate the uncertainty in the measured dimension d from the  
average pdf over 6 snapshots, we consider variations of the pdf
from snapshot to snapshot, which supposedly include both the intrinsic
temporal fluctuations of the pdf (if they exist) and the measurement
uncertainty in each snapshot. We think the measurement uncertainty at
small scales is probably small because of the large number
of sampling cubes (thus we put more weight to the pdf towards smaller
scales) and the snapshot-to-snapshot variations at these scales
may primarily reflect the intrinsic temporal fluctuations (which
corresponds to the variations in the predicted temperature from time
to time). We calculated the standard deviation of log (P) in different
$\epsilon$ bins over the 6 snapshots. In Fig. 5,  we plot the 1-$\sigma$ error 
bars around the data points for the pdf at $l=1/256 L$, which are normalized 
by the log-Poisson pdf with $d=1.64$. Error bars of similar size are also 
obtained for other scales (the size of error bars for the scale, L/64, is 
about 60\% larger than those in Fig. 5. This is because the measurement 
uncertainty is larger due to the smaller number of sampling cubes.). 
We see that the snapshot-to-snapshot variation is quite small except 
at the far right tail (the effect of which will be discussed later in \S 4.3). 
From the two curves representing the log-Poisson model with 
$d=1.56$ and $d=1.71$ (normalized to that with $d=1.64)$, we find that the 
model gives acceptable fits to the measured pdf within the 1-$\sigma$ error bars, 
for a range of values of the dimension $d$ between $d=1.56$ and $d=1.71$. 
We will evaluate the effect of this range of $d$ in our applications.

\begin{figure}
\epsscale{1.0}
\plotone{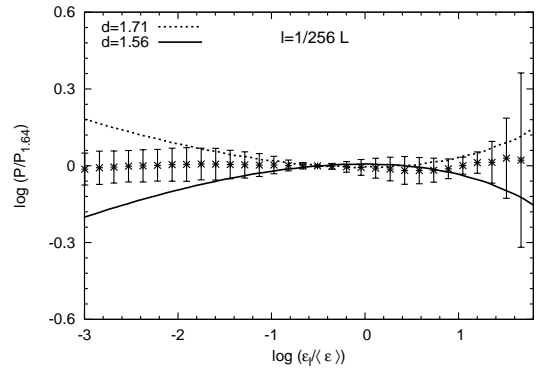}
\caption{Snapshot-to-snapshot variations of the dissipation rate pdf at $l=L/256$ in the simulation. The data points 
are the average pdf at $L/256$ (the same asterisk data points in Fig. 4) 
normalized to the log-Poisson pdf eq.~(\ref{eq9}) with the best-fit dimension, $d=1.64$. 
The error bars indicate the standard deviation (1-$\sigma$) of the pdf variations over the 6 snapshots. 
The error bars include both temporal variations of the pdf (if they exist) and measurement uncertainties. 
Also shown are the log-Poisson pdfs with $d=1.56$ and $d=1.71$, also normalized to that with $d=1.64$. 
The predicted distribution from the model with $d$ in the range (1.56 - 1.71) generally falls within 
the $1-\sigma$ error bars.                         
\label{f5}}
\end{figure} 

We have shown that the log-Poisson model agrees very well with the pdf from  resolved scales in our simulations for supersonic MHD 
turbulence.  As mentioned earlier, we are interested the dissipation rate pdf at the Kolmogorov scale, $\eta$, for applications 
to turbulent heating, and we assumed that the pdf at the scale $\eta$ can be obtained by extrapolating the log-Poisson model, 
proposed for the inertial range only, to the dissipation scale. We find that the assumption is well supported by our simulation 
result that pdf in the dissipation scales, such as $1/512$ $L$, and the resolution scale, is very close to that extrapolated from 
larger, inertial range scales, such as 1/128 $L$ and 1/256 $L$.

Reynolds numbers in interstellar turbulence are much larger than can be resolved by 
current numerical simulations. Because the intermittency theory can only be verified
by simulations with limited resolution, the extrapolation of the theory to a realistic 
value of $Re$ remains uncertain. Numerical experiments with much larger resolution are
required to test if the parameters of the theory, such as the dimension $d$, are 
independent of $Re$. However, based on the usual assumption of turbulence theory 
that the statistics in the inertial range are universal, we expect that a
theoretical model validated by our simulation, which resolves at least a small range 
of inertial scales, can be applied to turbulent flows with higher $Re$. If so, 
the statistics of the dissipation rate extracted from the inertial range of 
our simulation, and the derived dimension $d$ of the log-Poisson model, are universal.


We will apply the log-Poisson model to different interstellar clouds using the appropriate values of $Re$. For each type of clouds, 
we calculate the dissipation scale, $\eta$, from the Reynolds number, and use the pdf, eq.~(\ref{eq9}), at $l=\eta$, for the distribution 
of the turbulent heating rate.

\section{Energy Balance}

The turbulent heating rate per unit volume is given by, 
\begin{equation}
\Gamma_{\rm t}= n \mu m_H \epsilon({\bf x}, t)
\label{eq10}
\end{equation}
where $n$ and $m_H$ are the number density and the mass of the hydrogen atom, $\mu$ is the mean molecular weight 
(we adopt $\mu=2.35$ for molecular clouds and $\mu=1.23$ for atomic clouds), and $\epsilon({\bf x}, t)$ is the turbulent dissipation 
rate per unit mass discussed in \S~2. 
The heating rate is spatially inhomogeneous due to the intermittency of the turbulent dissipation. This gives rise to 
temperature fluctuations. We will use the distribution of $\epsilon$ given by eq.~(\ref{eq9}), and calculate the mass-weighted average temperature 
and the temperature probability distribution in dense molecular clouds and in cold diffuse atomic clouds.

We first consider conventional heating processes in these clouds, and then compare their effect with that of turbulent heating. 
Cosmic rays have been thought to be the primary heating source for molecular clouds. 
The heating rate by cosmic rays depends on the ionization rate, which is very uncertain (e.g., Caselli et al. 1998). The measured 
ionization rate in massive protostellar envelopes, where photoionization is insignificant, is about $2.6 \times 10^{-17}$ s$^{-1}$ 
(Van der Tak and Van Dishoeck 2000). Assuming that the energy input per ionization is $20$~ev (Goldsmith 2001), 
we take the cosmic-ray heating rate per unit volume to be,   
\begin{equation}
\Gamma_{\rm cr}=0.8 \times 10^{-27} n \hspace{1mm} {\rm ergs} \hspace{1mm} {\rm cm}^{-3} \hspace{1mm} {\rm s}^{-1} 
\label{eq11}
\end{equation} 
where the number density $n$ is in units of cm$^{-3}$.  

In the neutral atomic medium, the dominant heating source is the grain photoelectric heating. We use the photoelectric heating rate 
given in Wolfire et al. (2003), 
\begin{equation}
\Gamma_{\rm pe}= 1.3 \times 10^ {-24} n \epsilon_{\rm pe} G_0 \hspace{1mm} {\rm ergs} \hspace{1mm} {\rm cm}^{-3} \hspace{1mm} {\rm s}^{-1} 
\label{eq12}
\end{equation} 
where $G_0$ is the FUV flux in units of the integrated field of Habing (1968) and $\epsilon_{\rm pe}$ is the heating efficiency. 
Following Wolfire et al., we take $G_0 =1.7$ for the local ISM. For the typical cold neutral cloud temperature of 100~K, density of 50~cm$^{-3}$, 
and electron fraction of $10^{-4}$ (see Fig.~10 or eq.~(C15) in Wolfire et al. 2003), 
we get $\epsilon_{\rm pe}=1.4 \times 10^{-2}$ (using eq.~(20) in Wolfire et al. 2003), and hence
$\Gamma_{\rm pe} = 3.1 \times 10^{-26} n$ ergs cm$^{-3}$ s$^{-1}$. 

For molecular clouds, we take the cooling rate per unit volume, $\Lambda_{\rm g}$, from Table~2 of Goldsmith (2001), 
which gives coefficients and indices for power-law fits to the temperature dependence of the cooling rates at densities from 10$^2$ 
to 10$^7$~cm$^{-3}$. We use a linear interpolation to get coefficients and indices for densities not given in the table. 
The table is only for velocity gradient of 1~km~s$^{-1}$~pc$^{-1}$. According to eqs (2), (3) and (4) 
in Goldsmith and Langer (1976), the cooling rate per particle depends on the velocity gradient, $dv/dr$, 
and on the density only through their product, $n/(dv/dr)$. Therefore, Table 2 of Goldsmith (2001) can be converted to one for 
the cooling rate per particle as a function of $n(dv/dr)$. We then obtain the cooling rate per particle for a
given density and velocity gradient in the range of $n/(dv/dr)$ corresponding to the density range in the table.            

For cold neutral HI clouds, the dominant cooling process is the line cooling by the CII 158 $\mu$m fine structure transition.   
Considering the low electron fraction (10$^{-4}$) in these clouds, we neglect the contribution to the CII cooling from 
collisional excitation by electrons.
In \S~4.3, we compute the fraction of gas in cold diffuse clouds with high enough temperature ($\gsim 1000$~K) 
to activate the chemical reactions producing CH$^+$. We therefore include cooling by the OI 63 $\mu$m line, which 
becomes more important than CII cooling at $T>1000$~K. We adopt the cooling rates for CII and OI lines from Wolfire et al (2003). 
Note that this cooling rate is smaller than that given by Dalgarno and McCray (1972) by a factor of a few. This is 
because Wolfire et al. (2003) adopted C and O abundances derived from observations of UV absorption lines by 
the interstellar gas over many lines of sight studied by HST/FUSE, 
which are smaller than the solar abundances used in Dalgarno and McCray (1972) by a factor of 2-3.     

In our calculations, we will consider energy exchange between gas and dust grains and solve the coupled energy balance 
equations for gas and dust grains simultaneously. For the energy transfer rate from gas to dust grains, $\Lambda_{\rm gd}$, 
we take the formula,  
\begin{equation}
\Lambda_{\rm gd}=0.7 \times 10^{-33} n^2 T^{1/2} (T - T_{\rm d}) \hspace{1mm} {\rm ergs} \hspace{1mm} {\rm cm}^{-3} \hspace{1mm} {\rm s}^{-1}   
\label{eq13}
\end{equation}  
from Goldsmith (2001), where $T$ and $T_{\rm d}$ are, respectively, the gas and the grain temperatures in units of K (see also Black 1987, 
where the rate is larger by a factor of a few). Eq.~(\ref{eq13}) assumes a dust to gas ratio of 0.01, grain size of $1.7 \times 10^{-5}$~cm, 
grain density of 2~g~cm$^{-3}$, and an accommodation coefficient of 0.3 for grains and H$_2$.   
If $T>T_d$, the gas heats the dust and $\Lambda_{\rm gd}$ is a cooling term for the gas; vice versa, if $T<T_{\rm d}$, $\Lambda_{\rm gd}$ is a 
heating term for the gas. 

Dust grains in molecular clouds are heated by the diffuse UV-visible-IR interstellar radiation. 
Following Goldsmith (2001), we adopt a flux 
of $5.3 \times 10^{-3}$~ergs~cm $^{-2}$~s$^{-1}$ for this radiation field. Using the above values for the dust to gas ratio and the 
grain cross section (calculated from the grain size and density), the dust heating rate by the diffuse interstellar radiation is,  
\begin{equation} 
\Gamma_{\rm d}= 3.9 \times 10^{-24} n \chi \hspace{1mm} {\rm ergs} \hspace{1mm} {\rm cm}^{-3} \hspace{1mm} {\rm s}^{-1}  
\label{eq14}
\end{equation}  
To account for the attenuation of the radiation field due to dust extinction, we have reduced the heating rate by a factor 
of $\chi$. We take $\chi$ to be $\sim 10^{-1}-10^{-2}$, corresponding to a visual distinction of 2-5 mag., typical of a 
molecular cloud of size 1~pc. The value of $\chi$ affects only the dust temperature in such a cloud (\S 4.1).       
For inner regions of high-mass star-forming cores, the radiation from newly formed stars is the primary heating source of dust grains 
(see \S~4.2). The heating rate from the stellar radiation gives a high dust temperature. In that case, collisions of gas molecules 
with dust grains could provide a significant energy source for the molecular gas through $\Lambda_{\rm gd}$. 
For dust heating in atomic clouds, we use eq.~(\ref{eq14}) with $\chi=1$.       

The dust cooling rate is calculated by integrating over the wavelength, $\lambda$, the absorption efficiency, $Q_{\lambda}$, 
times the Planck function (Hollenbach and McKee 1979). For $\lambda \lsim 100 \mu$m (which corresponds to $T_{\rm d} \lsim$140~K), 
the absorption efficiency decays as $\lambda^{-2}$ (Draine and Lee 1984; Ossenkolf and Henning 1994). Using 
$Q_{\lambda} \simeq 4.5 \times 10^{-5}$ at the reference wavelength of $\lambda =790 \mu$m 
(Goldsmith 2001; Ossenkolf and Henning 1994), a numerical evaluation of the integral gives, 
\begin{equation}
\Lambda_{\rm d}= 4 \times 10^{-31} n T_{\rm d}^6 \hspace{1mm} {\rm ergs} \hspace{1mm} {\rm cm}^{-3} \hspace{1mm} {\rm s}^{-1}
\label{eq15}
\end{equation}                     
(Note that this result of our numerical calculation is 60 times larger than that given in Goldsmith (2001)). 
In our calculations below, the dust temperature is smaller than 140 K, eq.~(\ref{eq15}) is thus valid for the temperature range of 
interest here.    

Putting together heating and cooling for gas and dust grains and their energy exchange, the energy balance equations for 
gas and dust grains are, respectively, 
\begin{equation}
\Gamma_{\rm t} + \Gamma_{\rm cr} +\Gamma_{\rm pe} -\Lambda_{\rm g} - \Lambda_{\rm gd}=0    
\label{eq16}
\end{equation}
and 
\begin{equation}
\Gamma_{\rm d} + \Lambda_{\rm gd} - \Lambda_{\rm d} =0           
\label{eq17}
\end{equation}
The energy balance assumption is here justified because the cooling timescale of interstellar clouds 
we consider in the next section ($\sim 10^4$~yr) is smaller than their dynamical timescale ($\sim 10^6$~yr). 
We will solve these two coupled equations numerically.

\section{Applications}

We apply the intermittent turbulent heating model to three different interstellar medium environments: Molecular 
clouds, high-mass star-forming cores, and cold diffuse HI clouds. 
The spatial variations in the turbulent heating rate, $\Gamma_{\rm t}$, given by eq.~(\ref{eq10}), may result in
large spatial fluctuations of the gas temperature. Given the distribution function of $\Gamma_{\rm t}$, we can calculate 
the temperature probability function by solving eqs.~(\ref{eq16}) and (\ref{eq17}). 
However, there is a complication in converting the mass-weighted distribution of the heating rate to the 
mass-weighted temperature distribution. The various terms in eqs.~(\ref{eq16}) and (\ref{eq17}) have different density 
dependence and thus an exact calculation of the mass-weighted temperature distribution requires the joint statistics of 
density and dissipation rate fluctuations, which is beyond the scope of this study. We will therefore solve the
energy balance equations using simply the average gas density. This is equivalent to assuming that 
the density field and the dissipation field are not correlated, which is supported by their
low correlation coefficient of -0.11 and by Figures~\ref{f2} and \ref{f3}. Although the temperature distribution is not 
mass-weighted in an exact way, the resulting temperature distribution and the average temperature from our 
calculations are mass-weighted in the sense that they are computed from the mass-weighted distribution of the dissipation rate.     
In the case of molecular clouds, the mass-weighting of the temperature distribution discussed 
in \S~4.1 is almost exact. The cooling rate in these clouds has an almost linear dependence on density 
(Goldsmith and Langer 1978) so, except for the negligible coupling term $\Lambda_{\rm gd}$, the density dependence 
in all the terms of eq.~(\ref{eq16}) is the same.

We point out that the nearly independence between the dissipation rate
(per unit mass or per particle) and the density does not imply that the
temperature and the density are decoupled. The gas temperature is 
calculated from the balance between heating and cooling.
The cooling rate per particle depends on the density and thus the  
resulting temperature is correlated with the density through the cooling rate. 
In this sense, the correlation is similar to the case with a uniform heating 
rate (as usually assumed for traditional heating sources, e.g., 
the cosmic-ray heating): denser regions tend to have a lower temperature 
than lower density ones. However, the intermittent
fluctuations in the turbulent heating rate may also give rise to significant 
scatter around this general trend. This scatter may be probed through 
high-resolution molecular emission line maps. In such maps, the observed scatter 
around the predicted correlation between the projected temperature and density (i.e.,  
averaged along the line of sight) would be smaller than that in 3D space, as  
can be seen by comparing Fig.~1 with Fig.~2.

\subsection{Molecular Clouds}

We assume the following Larson's relations for the 1D velocity dispersion and the number density in molecular clouds,   
\begin{equation}  
\sigma_{\rm v}= 1 (L/1\hspace{1.5mm} {\rm pc})^{0.4} \hspace{1.5mm} {\rm km} \hspace{1.5mm} {\rm s}^{-1}  
\label{eq18}
\end{equation}
and 
\begin{equation}
n=2000 (L/1\hspace{1.5mm} {\rm pc})^{-1} \hspace{1.5mm} {\rm cm}^{-3}
\label{eq19}
\end{equation} 
where $L$ is the size of the cloud. The average dissipation rate per unit mass is
${\bar \epsilon} = 1/2 (\sqrt{3} \sigma_{\rm v})^3/L = 0.84 \times 10^{-3} (L/$1 pc)$^{0.2}$~ergs~g$^{-1}$~s$^{-1}$, 
where we used the conversion from the 1D rms velocity, $\sigma_{\rm v}$, to the 3D rms velocity, $U$, i.e., $U=\sqrt{3} \sigma_v$.
We have assumed that the turbulence is driven at the length scale of the cloud 
(see Basu and Murali 2001, who find that driving at smaller scale, for example by stellar outflows, 
would result in a turbulent heating rate that gives a CO luminosity in excess of the observations).
Therefore, the average turbulent heating rate per unit volume, given by eq.~(\ref{eq10}), is, 
\begin{equation}
{\bar \Gamma}_{\rm t} = 3 \times 10^{-27} n (L/1\hspace{1.5mm} {\rm pc})^{0.2} \hspace{1mm} {\rm ergs} \hspace{1mm} {\rm cm}^{-3} 
\hspace{1mm} {\rm s}^{-1} 
\label{eq20}
\end{equation}
In reality, the turbulence may be driven at a scale much larger than the cloud size, as suggested by various studies of
the velocity scaling in molecular clouds (e.g., Ossenkopf and Mac Low 2002; Heyer and Brunt 2004; Padoan et al. 2006).
However, from eq.~(\ref{eq20}), $\Gamma_{\rm t}$ is almost independent of $L$, and therefore eq.~(\ref{eq20}) is accurate 
within a factor of a few. By adopting the cloud size as the driving scale, we make a conservative assumption, which may 
slightly underestimate the turbulent heating rate and the resulting gas temperature.

Comparing eq.~(\ref{eq20}) with eq.~(\ref{eq11}), we see that the average turbulent heating rate in molecular clouds is of the same order of 
the cosmic ray heating rate, and possibly a few times larger. This average turbulent heating alone could maintain a temperature of 
$\sim$10~K in molecular clouds, and thus the cosmic ray heating, whose rate is highly uncertain, is not even needed to explain the
observed gas temperatures. Neglecting the spatial fluctuations in the turbulent heating rate, and using the average heating rate 
from eq.~(\ref{eq20}) in eq.~(\ref{eq16}), and the dust heating rate from eq.~(\ref{eq14}) in eq.~(\ref{eq17}), we find that the 
turbulent heating alone gives a temperature of 17~K for a 1-pc cloud. 
Similar temperatures are obtained for clouds of size 0.1~pc and 10~pc, because the heating rate from eq.~(\ref{eq20}) depends  
very weakly on $L$. In the density range of these clouds (200-2$\times 10^4$), the coupling between gas and dust is week, so 
$\Lambda_{\rm gd}$ is negligible in comparison with the gas and dust heating rates, 
and can be neglected. 
The dust and the gas temperatures can be obtained simply from their own energy balance (and thus the gas 
temperature is independent of $\chi$ in eq.~(\ref{eq17})), and, from the interstellar UV-visible-IR radiation field, 
the dust grains achieve a temperature of 7~K if $\chi=10^{-2}$, or 10~K if $\chi=10^{-1}$.   

A realistic calculation for the gas temperature needs to take into account the intermittent distribution of 
the turbulent heating rate. Note that the cooling rate in molecular clouds depends on temperature quite sensitively; 
for example at $n=1000$~cm$^{-3}$, $\Lambda_{\rm g} \propto T^{2.4}$ (Goldsmith 2001). Therefore, given the average 
temperature, the presence of temperature fluctuations makes the overall cooling rate larger. Conversely, 
given the average heating rate (and thus the overall cooling rate), the presence of heating rate 
fluctuations (which give rise to temperature fluctuations) leads to an average temperature lower than from uniform heating.
The larger the fluctuation amplitude (for example from stronger intermittency in the heating rate), 
the smaller the average temperature. This effect can be significant because the log-Poisson distribution
of the heating rate, eq.~(\ref{eq9}), is very skewed toward low values (see Figure~\ref{f4}). In other words,
the turbulent heating rate is below its average value in a large fraction of the mass. Because the hotter gas cools 
faster than the cooler gas, the contribution to the average temperature is mainly from the low-dissipation-rate 
part of the distribution. 

\begin{figure}
\epsscale{1.0}
\plotone{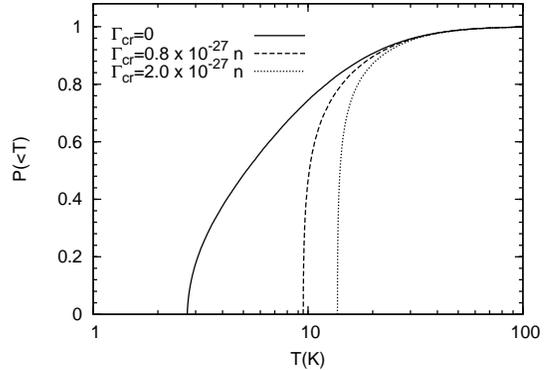}
\caption{Cumulative probability of temperature smaller than $T$ in a molecular cloud of size 1~pc.  The three curves correspond 
to the results for different cosmic ray heating rates. Without cosmic ray heating (solid line), the average temperature from turbulent 
heating is 8.5~K, and 20\% of the gas mass is essentially not heated, due to the intermittency of the turbulent heating rate. When 
included (dashed and dotted lines), the cosmic ray heating sets a lower limit for the temperature of 9.5~K and 14~K (corresponding 
to the uniform temperature in the absence of turbulent heating), for the two heating rates considered here. In these two cases, the 
average temperature is 13~K and 17~K.    
\label{f6}}
\end{figure}

To quantify this intermittency effect, we assume the distribution of the turbulent heating rate, $\Gamma_t$, follows the log-Poisson 
distribution, eq.~(\ref{eq9}), which we evaluate at the dissipation scale, $\eta$. The dissipation scale is $\eta = L Re^{-3/4}$, 
where the Reynolds number, $Re$, is defined as $Re=UL/\nu$. 
The kinematic viscosity is calculated as $\nu = c /(n \sigma)$ (see \S~2). We adopt a typical value for the cross section, 
$\sigma \sim 10^{-15}$~cm$^{2}$, and use the average temperature to calculate the sound speed, $c$. 
By numerically solving eq.~(\ref{eq16}) and eq.~(\ref{eq17}), we convert the distribution of $\Gamma_t$ into the temperature 
distribution. We assume the dimension of the intermittent dissipative structures to be $d=1.64$. The cumulative probability of the
gas temperature is plotted in Figure~\ref{f6} for a 1-pc cloud (solid line). 
We find that the average temperature of the distribution is 8.5~K. As expected, this is lower 
than 17~K, obtained using the average dissipation rate, but still a reasonable mean temperature for molecular clouds, in the
absence of cosmic ray heating.

We estimate the effect of the uncertainty in $d$ measured from
snapshot-to-snapshot variations, $1.56< d <1.71$, and find that this
range of values of $d$ corresponds to a range of 8.0-8.9 K for the  
average temperature, i.e.,  an uncertainty of only $\sim$ 6\% around the  
value of 8.5 K, corresponding to $d=1.64$. The effect on the average  
temperature is so small because the average temperature is
$\sim \langle \epsilon^{1/2-1/3} \rangle$ (since the cooling rate is
$\sim T^{2-3}$ for the density, $10^3-10^4$ cm$^{-3}$, in the molecular 
clouds of interest here), which is mainly contributed by the part of the pdf between
its peak and the average dissipation rate. From the error bars in  
Fig. 5, one can see there is little snapshot-to-snapshot variation 
in this part of the pdf. To further illustrate the effect of the degree 
of intermittency, we also calculated the average temperature using different 
values of the dimension of the dissipative structures.  For example, if $d=1$, 
as in the case of incompressible flows, the dissipation rate distribution would 
be less intermittent and the average temperature would be 11~K. On the other hand 
if $d=2$,  the average temperature would be lower, about 6~K.

Although the turbulent heating alone gives a reasonable average molecular cloud temperature of $\sim$ 8.5 K, 
a significant mass fraction of the clouds receives little heating from turbulent dissipation, because the 
distribution of the dissipation rate is skewed toward small values. 
For $d=1.64$, about 20\% of the mass is not significantly heated by turbulence and, in the absence of 
other heating processes, this mass fraction would be at the same temperature of 2.7~K as the cosmic microwave 
background (see Figure~\ref{f6}). The cosmic ray heating would of course dominate in regions with essentially 
no turbulent heating. In Figure~\ref{f6} we also show the temperature distribution including both turbulent and 
cosmic ray heating. To account for the uncertainty in the cosmic ray heating rate, we use two values, 
$\Gamma_{\rm cr}=0.8 \times 10^{-27}~n$~ergs~cm$^{-3}$ s$^{-1}$ (eq.~(\ref{eq11}); dashed line) and 
$\Gamma_{\rm cr}=2.0\times 10^{-27} n$~ergs~cm$^{-3}$~s$^{-1}$ (dotted line). The heating rate by cosmic rays sets the lower 
temperature limit (corresponding to the uniform temperature in the absence of turbulent heating).  
For $\Gamma_{\rm}= 0.8 \times 10^{-27}$~ergs~cm$^{-3}$~s$^{-1}$, this lower limit is 9.5~K.   
In this case, the average temperature is 13~K and most gas (80\% of the mass) has a temperature lower than the average.  
The larger cosmic ray heating rate chosen here, $\Gamma_{\rm cr}=2.0\times 10^{-27} n$~ergs~cm$^{-3}$~s$^{-1}$,  
gives a lower limit of 14 K and an average temperature of 17 K. The tail of large turbulent dissipation rate gives a 
small mass fraction of gas with high temperature. We find that only 1\% of the gas could be hotter than 40~K.   

We have shown that the turbulent heating by itself can maintain an average temperature of about 8.5~K in molecular clouds. 
Due to the intermittency of this heating process, a finite but small fraction of gas has temperature higher than 40~K, and 
a significant fraction of the gas is essentially not heated by turbulence, but rather by cosmic rays. 
In the presence of cosmic rays, turbulent heating increases the average temperature by a few degrees, from 9.5~K to 13~K 
for the cosmic ray heating rate given by eq. (\ref{eq11}).

Similar calculations can be done for dense dark cloud cores. 
We use the dense cores in the Orion A molecular cloud complex as an example
(e.g., Li et al. 2003, 2007), and focus on the ones with no apparent internal 
sources for simplicity\footnotemark\footnotetext{In dense cores 
containing young stellar objects (e.g., those discussed in Johnstone and Bally 2006), 
one needs to consider heating from the central sources and the calculation would be
similar to 
that for the high-mass star-forming cores to be discussed in \S 4.2}. 
The sizes of these cores are typically 0.05-0.2 pc (e.g., Tatematsu
et al. 1993, Li et al. 2003, Ikeda et al. 2007) and the typical average
density is $10^4-10^6$ cm$^{-3}$. Observations using
tracers such as CS (e.g., Tatematsu et al. 1993), NH$_3$ (e.g., Li et al. 2003), 
H$^{13}$CO$^+$ (Ikeda et al. 2007), and other molecular transitions show that the
spectral lines have FWHM (full width at half maximum) in the range 0.5-2 km s$^{-1}$, 
corresponding to 0.22-0.89 km s$^{-1}$ for the 1D rms velocity, 
$\sigma_{\rm v}$. We emphasize that we evaluate the dissipation rate based on 
the rms velocity, while observational papers usually  
give the FWHM. We obtain the 1D rms velocity from the 
conversion $\sigma_ {\rm v}=$ FWHM/2.355.

We find that the typical values of $\sigma_{\rm v}=0.5$ km s$^{-1}$ and  
$L=0.1$ pc follow quite closely Larson's relation given in
eq.~(\ref{eq18}), implying that the turbulent heating rate is similar  
to that in typical regions of molecular clouds discussed above. 
A complication in evaluating the temperatures in these cores is the 
possible presence of external UV sources (e.g., from the Trapezium star 
cluster, see Li et al. 2003). We also need to consider the inner and 
outer regions of these cores separately. The central region is 
self-shielded from the diffuse interstellar radiation field or 
the external UV sources, and thus has a lower dust temperature than the 
outer regions if there are no internal sources. 
Furthermore, the gas and dust temperatures are well coupled in the central region, 
due to the high density ($\sim 10^6$ cm$^{-3}$) (see \S 4.2 for a more 
detailed discussion on the gas-dust coupling). We find that including 
turbulent heating has little effect to the coupled gas and dust 
temperature in the central region and the temperature there is 
essentially determined by the dust heating rate from the diffuse interstellar 
radiation field and the external UV sources. 

At the lower density outer edge, gas and dust are thermally 
decoupled, and we find that the effect of turbulent heating to the gas 
temperature is similar to that in the molecular clouds discussed 
above. As an example, we assume that the dust heating is from the diffuse 
interstellar radiation field and the rate is given by eq.~(\ref{eq14}), 
with $\chi \sim1$ at the edge. This rate gives a dust temperature of 15 K,  
close to the average dust temperature of 17 K found in the observations 
by Lis et al. (1998). Assuming the gas density is $10^4$ cm$^{-3}$ at 
the edge, our model with $d=1.64$ shows that turbulent dissipation 
alone can heat the gas to an average temperature of 11 K, close to the 
temperature of 10 K due to cosmic-ray heating alone. Combining the comic-ray heating
with the turbulent heating results in a gas temperature of 14 K, which is in 
general agreement with the observed temperature (e.g, Li et al. 2003). 
The effect of turbulent heating here is thus the same as that in molecular 
clouds studied above. Some cases of higher gas temperature have also been found in 
dense cores in Orion, which can be explained by external UV sources 
(or internal sources if they are present). UV sources directly heats 
dust grains and increase the dust temperature. Since in the denser inner regions 
gas and dust are thermally coupled, the gas temperature there may be increased 
to the observed higher values by UV sources through energy transfer with dust grains. 
(The gas temperature at the outer edge with a density $\sim 10^{-4}$ cm$^{-3}$ 
may not increase considerably because the gas and dust temperatures are there
decoupled.)   

In summary, turbulent heating in central regions of the dense cores 
in Orion have negligible effects to both gas and dust temperatures, while 
its role for the outer regions is similar to that in the molecular clouds 
following Larson's relation. Turbulent heating alone can heat the gas to a
temperature of 11 K, and a few degrees higher in combination with cosmic ray
heating. The effect of turbulent heating is more prominent in objects with
much larger turbulent intensity, such as in the high-mass star-forming  
cores to be discussed in \S~4.2, whose average temperature is 
significantly increased, or in places where the intermittent tail 
of the heating rate pdf plays an important role for the chemistry, 
such as in HI clouds, which we study in \S~4.3.

\subsection{High-Mass Star-Forming Cores}

We now apply the turbulent heating model to the high-mass star-forming
cores observed in various CS transitions by Plume et al. (1997).
These cores were originally selected by the presence of water masers,
suggesting the formation of massive stars, as also shown by their
large FIR luminosity (Mueller et al. 2002). The characteristic mean
density of these cores is $10^6$~cm$^{-3}$, the mean size 0.3~pc  
(average values over the objects listed in Plume et al. (1997) and Shirley et al. (2003) ).
These cores are very massive, with a typical mass of $\sim 1000 M_\odot$
(Plume et al. 1997, Shirley et al. 2003) (much more massive than the dense cores 
in Orion discussed in \S 4.1). We are interested in these cores because of their extremely large  
turbulent intensity. The observed FWHM is in the range of 2-12 km s$^{-1}$ (Shirley et al.  
2003) with an average of $\sim$ 6 km s$^{-1}$. This average correspond to a  
1D rms velocity of 2.5~km~s$^{-1}$. 
The average turbulent dissipation rate in these dense cores is very large, $\bar \epsilon = 4.4 \times 10^{-2} $~ergs~g$^{-1}$~s$^{-1}$, 
approximately 50 times larger than that in a molecular cloud following Larson's relations (see \S~4.1). However, 
this heating rate per unit mass, which can be written as $\sim 0.02~L_\odot/M_\odot$, is 
much smaller than the average (FIR) bolometric luminosity to mass ratio, 140~$L_\odot/M_\odot$, found by Mueller et al. (2002). 
Clearly, turbulent heating in these cores could not account for even a tiny fraction of the FIR radiation.   
The FIR emission must be primarily due to the processed UV photons from the central massive stars, which directly heat the dust grains, 
resulting in a fairly high dust temperature. The primary gas heating source in these cores is thought to be the energy exchange 
with dust grains by collisions (Evans 1999). We want to investigate if the turbulent heating may provide a significant 
energy source for the gas in these dense cores and to estimate the effect of the turbulent heating on the gas temperature. 

The gas density and the dust heating rate from the central stellar sources (which can be estimated from the dust temperature) are needed to 
calculate the gas temperature from eqs.~(\ref{eq16}) and (\ref{eq17}). The gas density, through the collision frequency, plays a crucial role in 
determining how well gas and dust grains are thermally coupled (see eq.~(\ref{eq13})). The degree of coupling decreases with the 
distance to the center, $r$, as the density decreases. The density profile in these cores is usually assumed to be a power-law, 
$n(r) \propto r^{-\alpha}$, where $\alpha$ is in the range 0.5-2.5 (Mueller et al. 2002). In this range of $\alpha$, most of the
gas mass is in the outer regions at large $r$; even for the steepest slope, $\alpha=2.5$, about 70\% of the gas mass is outside 
$r=0.1$~pc, for a core of size 0.3~pc. Since a large fraction of the gas mass is in the outer regions, one may expect that the density 
there is about the average density, $10^6$~cm$^{-3}$, given earlier. However, the measured average density of $10^6$~cm$^{-3}$ is 
evaluated from radiative transfer modelling of the emission lines of tracer molecules (Shirley et al. 2003), and thus probably 
reflects the average density along the line of sight, which is dominated by inner regions for a density profile steeper 
than $r^{-1}$. In that case, the density at the outer regions is much smaller than $10^6$ cm$^{-3}$.\footnotemark\footnotetext{Plume et al. (1997) 
found that the average density, $\sim 10^6$ cm$^{-3}$, derived from their radiative transfer model, gave a core mass larger 
than the virial mass by more than an order of magnitude. They realized that the mass discrepancy might have originated from 
the assumption of uniform density in their model. This suggests that the true average density, determined mainly by the outer regions, 
is smaller than $10^6$~cm$^{-3}$ by one or two orders of magnitude, supporting our choice of $10^4$-$10^5$~cm$^{-3}$.} 
For example, if $\alpha=2$, normalizing the line-of-sight average density to $10^6$~cm$^{-3}$ and assuming that the radius 
at the inner edge of the core is 100~AU, we find that the density at 0.1~pc is $10^4$~cm$^{-3}$
(while the density at 0.01~pc is close to the measured average density of $10^6$~cm$^{-3}$).  
To account for the possibility of a flatter density profile, we take $n$ at 0.1~pc to be $10^4-10^5$~cm$^{-3}$. 
We did similar estimates for the density at inner regions and found that, almost independent of the slope, $\alpha$, in the 
range 0.5-2.5, the gas density at 0.01~pc is always close to the measured average density of $10^6$ cm$^{-3}$.

Our calculations show that 10$^6$~cm$^{-3}$ is a critical density 
for the gas and dust temperatures to be well coupled. Thus the gas temperature behavior within 0.01~pc is different 
from that outside 0.01~pc. In our calculations, we will refer to two characteristic values of the radius, $r=0.01$~pc and 
$r=0.1$~pc, representative of the different gas temperature behaviors in the inner and outer regions of these cores.      

The dust temperature depends on the heating rate from the central stellar source, which decreases with increasing radius
roughly as $r^{-2}$. Using the cooling rate given by eq.~(\ref{eq15}), we obtain $T_{\rm d} \propto r^{-1/3}$, which is 
close to the power-law fit ($T_{\rm d} \propto r^{-0.4}$) to the numerical result from the radiative transfer model in 
Mueller et al. (2002) (the difference in the power-law indices here may suggest that a radiative transfer calculation 
gives a dust heating rate decreasing faster than $r^{-2}$, or a dust cooling rate increasing with $T_{\rm d}$ more slowly 
than $\propto T_{\rm d}^6$). We find that the energy transfer from dust 
to gas or from gas to dust (which is possible in the presence of turbulent heating of the gas) is negligible in comparison 
with the dust heating rate from the stellar photons, so the dust temperature is completely determined by the central stellar source.
For a typical bolometric luminosity of $10^4$~$L_\odot$, the dust temperature decreases from about 65~K at $r=0.01$~pc, 
to about 25~K at $r=0.1$~pc (see Fig. 9 in Mueller et al. (2002)). 

We now calculate the gas temperatures at 0.01~pc and 0.1~pc.  We first study the inner region using $T_{\rm d}=65$~K, 
and a density of $10^6$~cm$^{-3}$. This dust temperature implies a heating rate of $3 \times 10^{-14}$~ergs~cm$^{-3}$~s$^{-1}$ 
by UV photons from the central stars (estimated from the dust cooling rate, eq.~(\ref{eq15})). Fixing this dust heating rate, 
we first calculate the gas temperature assuming the gas is heated only 
by collisions with dust grains. By solving eqs.~(\ref{eq16}) and (\ref{eq17}), we find a gas temperature of 57~K, suggesting a strong 
thermal coupling between gas and dust, as mentioned above. The gas heating rate per unit volume by dust grains is 
$3.7 \times 10^{-20}$~ergs~cm$^{-3}$~s$^{-1}$, which is only $\sim 10^{-5}$ times the heating rate of the dust grains 
by the stellar radiation field. Thus the dust temperature is not affected by the energy transfer to the gas. 
We then calculate the gas temperature at $r=0.01$~pc assuming the gas is heated both by the dust and by the turbulence.
The average turbulent heating rate in these cores is $1.3 \times 10^{-19}$~ergs~cm$^{-3}$~s$^{-1}$, which is larger than the rate 
from collisions with dust grains when the turbulence is neglected. 
The inclusion of turbulent heating should therefore result in a gas temperature larger than 57~K. Using again the intermittency model 
with $d=1.64$, we obtain a mean gas temperature of 64~K, very close to the dust temperature of 65~K. 
In summary, the gas and the dust temperatures in this dense inner region are very well coupled, but the turbulent heating can still
increase the gas temperature by a few degrees. 

The outer region is interesting because it contains most of the mass of the core and because its density is lower and so
gas and dust are less well coupled.  We carry out the same calculation as above for the outer region, at $r=0.1$~pc, assuming a 
density of $n=10^4$-$10^5$~cm$^{-3}$ and a dust temperature of $T_d=25$~K. Neglecting turbulent heating, we obtain
a gas temperature of 17~K and 6~K, for $n=10^5$ and $10^4$~cm$^{-3}$ respectively
(including cosmic ray-heating increases the temperature to 18~K and 10~K respectively). In this case the gas temperature is 
considerably smaller than the dust temperature due to the weak coupling (while with $n=10^6$ cm$^{-3}$ the gas and dust
temperatures would be almost equal). As we did for the inner region, we now include the turbulent heating. For illustration, 
we assume the average dissipation rate is independent of radius. Without energy transfer between dust and gas, we find that 
the intermittent turbulent heating alone would give a temperature of 36~K and 35~K, for $n=10^4$ and $10^5$~cm$^{-3}$
respectively. The gas temperature is here nearly independent of the density because at densities above $10^4$~cm$^{-3}$
the cooling rate per particle is almost constant (see Goldsmith and Langer 1978). Because in this outer region the thermal
coupling between dust and gas is weak, we find that including both the energy transfer between dust and gas and the intermittent
turbulent heating, the gas temperature is still 35-36~K. In other words, in the outer region of the cores, containing most of the core mass,
the gas is thermally decoupled from the dust and its temperature is completely determined by the turbulent heating. 
Similar to the case of molecular clouds, we find that range of $d$ from
1.56 to 1.71 results in a small range for the average temperature, within 1.5-2 K
(i.e., $\sim$ 5\%), around the value 35-36 K obtained for $d=1.64$.

Our result is in general agreement with the observations of high-mass  
protostellar objects by Leurini et al. (2007).  By fitting the observed methanol  
spectral lines, they derive the physical parameters of a model consisting of 
an inner core, an extended component, and an outflow. In particular, 
they find the kinetic temperature is in range of 22-40 K in the extended component (corresponding to  
the outer regions in our model) where the FWHM linewidth is 2.8-4.2 km $^{-1}$  
(see their Table 5) over an average size of 0.1 pc (smaller then the value  
of 0.3 pc we adopted above). With these specific parameters, and the density  
range of $10^5-10^6$ cm$^{-3}$ (Table 5 in Leurini et al. 2007), our model would
predict a temperature range of 28-39 K. Although the velocity dispersion
here is lower than adopted in our example, the size is also smaller,  
resulting in a dissipation rate and a temperature comparable to those in our  
example above. Future observations with high spatial resolution will provide a  
better test of our model.

We conclude that, although negligible in comparison with the dust heating rate by the central stellar sources, 
turbulent heating in high-mass star-forming cores provides an important energy source for the gas. 
For a characteristic density of $10^6$~cm$^{-3}$ in the inner regions, the gas and dust temperatures are well coupled. 
The turbulent heating increases the gas temperature only by a few degrees. This slight temperature increase in the inner 
regions may not have important observational or dynamical consequences. On the other hand, the turbulent heating has 
significant effects in the outer regions, where the density is lower and the coupling of dust and gas is weaker than in the 
inner regions. For a characteristic dust temperature of 20-30~K and a density of $10^{-4}$-$10^{-5}$, the 
turbulent heating increases the gas temperature from less than 10-20~K (lower than the dust temperature) up to about 36~K. 
This large increase in the gas temperature by a factor of 2-3 due to turbulent heating, in the region containing most of the core mass, may have 
important effects on the evolution of the cores, on their star formation efficiency, and on their stellar initial mass function.
Future observations of these cores may help determine their gas temperature with sufficient accuracy to 
constrain the relative importance of turbulent heating.

\subsection{Cold Diffuse HI Clouds}

The observed abundance of CH$^+$, HCO$^+$, and OH molecules in cold diffuse clouds (Crane et al. 1995, Gredel 1997, Lucas and Liszt 1996, 
Liszt and Lucas 1996) 
suggests the existence of hot regions inside these clouds. The activation temperature of reactions producing these molecules is $>1000$~K,
while the cloud average temperature is $<100$~K. Dissipative heating 
by MHD shocks and vortex tubes have been proposed to explain the existence of hot regions (see, e.g., Pety and Falgarone 2000). 
These dissipative structures are the most intermittent structures in the log-Poisson  model of the turbulent  
energy dissipation. Since the model gives the distribution of the dissipation rate at all intensity levels, 
it can be used to calculate the cumulative probability corresponding to the mass fraction of regions with $T>1000$ K. 
Falgarone and Puget (1995) performed a similar calculation. They used the probability distribution of the 
velocity difference from experiments with incompressible flows, where the strong dissipative structures 
responsible for the hot regions are vortex tubes. It is not clear whether their result applies to supersonic turbulence 
in diffuse neutral clouds, where both shocks and vortex tubes contribute to high temperatures. 
The log-Poisson intermittency model for supersonic turbulence includes both types of dissipation structures. 
We will compare the result by Falgarone and Puget (1995) with that from the log-Poisson model for incompressible turbulence
and for supersonic flows presented in \S~2.                    

For the characteristic density and length scale of cold neutral clouds we adopt $n=50$~cm$^{-3}$ and $L$=1~pc (Helies and Troland 2003). 
The observed 1D rms turbulent velocity, $\sigma_{\rm v}$, at the scale of 1~pc, is between 1 and 3~km~s$^{-1}$ (Hennebelle et al. 2007). 
The average turbulent heating rate is $1.5 \times 10^{-27} ( \sigma_{\rm v}/1$ km s$^{-1})^3 $~ergs~s$^{-1}$ per H atom. 
We will treat $\sigma_{\rm v}$ as a parameter and study the dependence of the mass fraction of hot regions on $\sigma_{\rm v}$. 
As discussed in \S~3, the photoelectric heating rate in cold neutral clouds is $3.1 \times 10^{-26}$ ergs s$^{-1}$ per H atom. Neglecting 
turbulent heating, this gives a temperature of 53 K using the CII and OI cooling rates from Wolfire et al. (2003). 
At $n=50$ cm$^{-3}$, the gas-dust coupling is very weak and their energy balance equations can be solved separately. 

We now consider the effect of turbulent heating. First we neglect intermittency and use the average dissipation rate, and then
we include the effect of intermittency using the log-Poisson model. Assuming
$\sigma_{\rm v}=2$~km~s$^{-1}$ as an example, we find that the temperature increases to 63~K, much smaller than the
1000~K required for production of CH$^+$. Temperatures much higher than the average can only be obtained
if intermittency is included. In the temperature range from 100~K to 8000~K, the cooling rate as a function of temperature is very flat, 
meaning that hotter gas is more difficult to cool. This suggests that temperature fluctuations in this range of values 
reduce the overall cooling efficiency. Therefore, the net effect of fluctuations in the dissipation rate is that of 
producing a higher mean temperature than in the case of a uniform dissipation rate (contrary to the case of molecular 
clouds, where the cooling rate depends on temperature very sensitively). For example, if the dimension of the most intense
dissipative structures is $d=1.64$, the average temperature derived for $\sigma_{\rm v}=2$ ~km~s$^{-1}$ is 100~K (versus 63~K in the
case of a uniform dissipation rate). Furthermore, the flat cooling curve helps the generation of hot regions due to
the fluctuations in the dissipation rate, so the extended high-rate tail of the dissipation rate  
distribution translates into a fairly extended high temperature tail in the temperature distribution. 
 
\begin{figure}
\epsscale{1.0}
\plotone{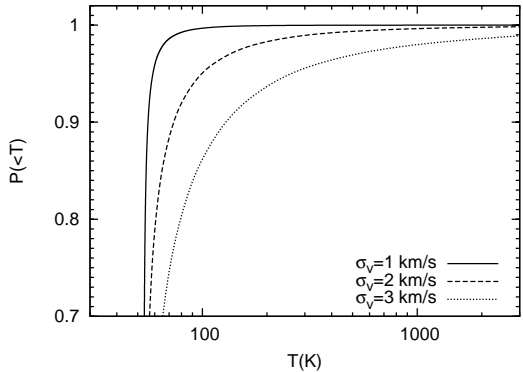}
\caption{Cumulative probability of regions with temperature smaller than $T$ in diffuse HI clouds. The three curves   
correspond to the results for three different values of the rms turbulent velocity, $\sigma_{\rm v}=1$, 2, and 3~km~s$^{-1}$,
at the scale of 1~pc. The mass fraction of gas with $T>1000$~K for these three cases is 3.6$\times 10^{-5}$, $3 \times 10^{-3}$, 
and 0.02, respectively. The observed CH$^{+}$ abundance requires that $\sigma_{\rm v}$ is at least 3~km~s$^{-1}$ at 1~pc. 
Most of the gas has a temperature below 100 K.    
\label{f7}}
\end{figure}  

We calculated the gas temperature probability distribution using $d=1.64$ for the most dissipative structures in 
the intermittency model. Figure~\ref{f7} shows the cumulative probability of temperatures lower than $T$ for different rms velocities.  
From the cumulative probability curve, we can read the fraction of gas with temperature larger than 1000~K. 
For $\sigma_{\rm v}=1$, 2, and 3~km~s$^{-1}$, the probabilities of finding $T>1000$~K are 3.6$\times 10^{-5}$, 3$\times 10^{-3}$ 
and $0.019$, respectively. Since the hot gas fraction needed to explain the observed abundance of CH$^+$ molecules is a few percent 
(Gredel et al. 1993), our result shows that cold neutral clouds, where molecules such as CH$^+$ are detected, must have 
a minimum rms turbulent velocity of 3~km~s$^{-1}$ at the scale of 1~pc\footnotemark\footnotetext
{We note that Sheffer et al. (2008) find a similar minimum velocity dispersion from a different model proposed by 
Federman et al. (1996) to explain the observed CH$^{+}$ abundance and its decoupling from the OH abundance. 
In their model, reactions between neutral and ionic molecules, e.g., the production of CH$^+$ from C$^+$ and H$_2$, 
are accelerated by the neutral-ion velocity, which is assumed to be the amplitude of MHD waves and is 
accounted for by using a non-thermal exponent for the effective temperature. We will investigate the intermittent distribution of 
the neutral-ion velocity in MHD turbulence and its effect on chemistry in HI clouds in a future work.}.
The uncertainty in $d$ within the range 1.56-1.71 corresponds to a range of the cumulative probability of only 
$\sim$ 2-3\% around the value of 0.019 found for $d=1.64$ and $\sigma_{\rm v}=3$ km s$^{-1}$.
Therefore, the range of $d$ does not affect our conclusion for the  
minimum rms velocity required to produce enough hot regions to explain the abundance of CH$^+$ molecules.
One may expect the large uncertainty at the far right tail (Fig. 5) may
significantly affect the cumulative probability. However, for $\sigma_{\rm v}= 3$ km s$^{-1}$, 
1000 K corresponds to a dissipation rate only 10 times  
larger than the average (see below for a discussion of a wider range of $d$, between  
1 and 2 ). Therefore the cumulative probability is not determined by the far  
tail, but mainly by the part of the pdf where there are no significant snapshot-to- 
snapshot variations, as can be seen from Fig.~\ref{f5}.
Figure~\ref{f7} also shows that most of the gas has temperature 
below 100~K, which is the mean temperature in the case of $\sigma_{\rm v}=2$~km~s$^{-1}$. This is because the skewness of the dissipation rate 
distribution favors values below the average. 
       
To study how the degree of intermittency affects the mass fraction of hot regions, we consider different values (beyond the range 1.56-1.71) 
of the dimension of the most intense 
dissipative structures. We choose two special vales, $d=1$ and $d=2$, corresponding to vortex tubes and shocks. 
The dissipation rate is more intermittent for larger values of $d$. We find that, if $\sigma_{\rm v}=1$~km~s$^{-1}$, the cumulative 
probability for $T>1000$ K depends on $d$ very sensitively. It is $2 \times 10^{-6} $ and $2 \times 10^{-4}$, respectively, for $d=1$ and $d=2$.  
However, for $\sigma_{\rm v}=3$~km~s$^{-1}$, the degree of intermittency does not affect the probability by much; the cumulative probability 
is 0.015 and 0.021 for $d=1$ and $d=2$, close to the value of 0.019 that we previously found for $d=1.64$. 
This is because for $\sigma_{\rm v}=1$~km~s$^{-1}$ the average 
dissipation rate is low, and a temperature of $T=1000$~K requires the far tail of the dissipation rate distribution, 
which strongly depends on the degree of intermittency. For $\sigma_{\rm v}=3$~km~s$^{-1}$, instead, $T=1000$~K corresponds to a 
dissipation rate only approximately 10 times larger than the average. In the range of values about an order of magnitude around the average,
there is no considerable difference in the dissipation rate distribution for different dimensions of the dissipative structures. 

Falgarone and Puget (1995) adopted an rms velocity of 3~km~s$^{-1}$ at 1~pc, and used experimental results for the probability distribution 
of the dissipation rate of incompressible turbulence (converted from the velocity difference distribution). Their model corresponds to 
our model with $d=1$. They obtained a mass fraction of a few percent (see their Figure 4) for gas with $T>1000$~K, 
consistent with our result using the same parameters. Therefore, Puget and Falgarone's model, physically valid only if the turbulence in 
cold atomic clouds were subsonic, gives by chance a result similar to that of our model for supersonic turbulence.     

In summary, we have shown that the intermittent turbulent heating can generate enough hot regions to produce the observed 
CH$^+$ abundance in cold neutral clouds if the turbulent rms velocity, $\sigma_{\rm v}$, is large enough. The required minimum value 
for $\sigma_{\rm v}$ is approximately 3~km~s$^{-1}$ at a length scale of 1~pc.

\section{Discussion and Conclusions}

We have studied the energy dissipation and the heating in supersonic turbulence. The turbulent dissipation is characterized by 
strongly intermittent fluctuations. A significant fraction of the kinetic energy is viscously dissipated in the finest, most intermittent
structures, giving rise to a broad tail in the probability distribution function of the dissipation rate. 
To study the turbulent heating in interstellar clouds, a theoretical model is needed for the probability distribution of the dissipation 
rate at the dissipation scale, $\eta$. We have generalized the log-Poisson model, originally proposed for incompressible turbulence 
by She and Leveque (1994), to supersonic turbulence. Because the dissipation scale, $\eta$, cannot be resolved by current numerical 
simulations, we have used results from resolved inertial-range scales in our numerical simulation of supersonic and super-Alfv\'{e}nic 
turbulence as a guideline for the sub-grid scales. We have found that the log-Poisson model, with a fractal dimension $d=1.64$ for the 
most intermittent dissipative structures, gives an excellent fit to the mass-weighted probability distribution of the 
dissipation rate at resolved scales in the simulation. Extrapolating the model to the corresponding Reynolds 
numbers, we have studied the turbulent heating in molecular clouds, high-mass star-forming cores, and cold diffuse neutral clouds. 
Here we summarize our results. 
\begin{enumerate}  
\item 
In typical molecular clouds, the average turbulent heating rate exceeds the cosmic ray heating rate by a factor of 3-4. 
Fluctuations in the heating rate give a lower mean temperature than expected from the average heating rate; temperature 
fluctuations make cooling more efficient because in molecular clouds the cooling rate increases sensitively with 
temperature. Taking intermittency into account, the turbulent heating alone gives a mean temperature of approximately 8.5~K, 
close to the characteristic temperature of molecular clouds. This would suggest that cosmic rays are not even needed to explain the 
thermal balance in molecular clouds. However, due to the intermittent fluctuations in the turbulent heating rate, a significant mass 
fraction of the molecular gas is not heated by the turbulence. Cosmic ray heating would dominate in these regions. Assuming a 
cosmic ray heating rate of $0.8 \times 10^{-27} n \hspace{1mm} {\rm ergs} \hspace{1mm} {\rm cm}^{-3} \hspace{1mm} {\rm s}^{-1}$, 
the turbulent heating increases the average temperature by a few degrees, from 9.5~K to 13~K.
We also find that turbulent heating plays a similar role in the outer regions of the dark cloud cores in the 
Orion molecular cloud complex.   
 
\item 
Turbulent heating provides an important energy source for the molecular gas in high-mass star-forming cores. Assuming spherical 
symmetry, in the inner regions of these cores (within $\sim 0.01$~pc from the center), stellar sources heat the dust grains to a relatively 
high temperature. Due to the high density of these inner regions ($\sim 10^6$~cm$^{-3}$), the gas and dust are thermally 
coupled, and the gas temperature is close to the dust temperature even in the absence of turbulent heating. The turbulent heating 
increases the gas temperature only by a few K. On the other hand, in the outer regions ($\sim 0.1$~pc from the center), where most of
the core mass resides, the turbulent heating results in a considerable increase in the gas temperature. The low density of these regions
($10^4$-$10^5$~cm$^{-3}$) makes energy exchange between gas and dusts inefficient and, without turbulent heating, the energy 
transfer from the dust heats the gas only to 10-20~K. Inclusion of turbulent heating increases the temperature to approximately 36~K.
Because turbulent heating causes a large temperature increase in most of the core mass, it may have important implications for 
the dynamical evolution of the cores and for their star-formation process, and it may also be probed by future observations.

\item     
The intermittent turbulent heating in diffuse HI clouds can give rise to regions much hotter than the average temperature.  
These warm regions have been used to explain the existence of molecules such as CH$^+$ in HI clouds, whose production needs 
high temperatures. Using the log-Poisson intermittency model for supersonic turbulence, we find that a turbulent rms velocity 
of 3~km~s$^{-1}$ at 1~pc is sufficient to account for the observed abundance of these molecules, which extends
the earlier result, based on incompressible turbulence, by Falgarone and Puget (1995).

\end{enumerate} 

We point out that thermal conduction is neglected in our calculations. Conduction tends to transport thermal energy to fill in 
regions not significantly heated by dissipation, and thus may, to some degree, erase the fluctuations in the heating rate. 
As discussed in \S~4, more intermittent fluctuations in the heating rate give a smaller average temperature in molecular clouds. 
Therefore, if thermal conduction were included, the average temperature in molecular clouds and in the outer regions of 
high-mass star-forming cores would be even larger, making turbulent heating even more important in these two cases (\S~4.1 and \S~4.2). 
The situation is different in cold diffuse HI clouds (\S~4.3). If the fluctuations in the heating rate are less intermittent 
there, the average temperature becomes smaller and, more importantly, the tail of the temperature probability distribution, 
needed for CH$^{+}$ production, would be less extended. We estimate whether and how much thermal conduction would change our results 
by calculating the conduction length scale, $l_c \simeq \sqrt{\kappa t_c}$, during a cooling time scale, $t_c$, where $\kappa$ is the 
thermal conduction coefficient (approximately equal to the kinematic viscosity, $\nu$). This is the scale over which thermal conduction 
can homogenize before the heat from turbulent dissipation is radiated away. We find that in molecular clouds and in HI 
clouds $l_c$ is smaller than (but comparable to) the dissipation length scale, $\eta$. Therefore, the heat generated in 
the most intermittent structures cannot be transported far from these structures by thermal conduction. This justifies our choice of 
computing the temperature distribution by using the dissipation rate distribution evaluated at $\eta$. In the outer 
regions of high-mass star-forming cores, $l_c$ is about an order of magnitude larger than $\eta$ (the latter is very 
small because of the very large Reynolds number). Using the distribution of the dissipation rate at $l_c$ instead of $\eta$ 
(assuming thermal conduction homogenizes the temperature over a size of $l_c$) gives an average temperature of 40~K 
in the outer regions of these cores, a little higher than from the distribution of the dissipation rate evaluated at $\eta$. 
However, the conduction process in the presence of turbulent motions is more complex than described by the 
above estimate, and a detailed study of its effects is beyond the scope of this paper. 

Although so far neglected, the process of turbulent heating may play an important role in the process of star formation. 
The mean temperature in molecular clouds defines the mean Jeans mass, which may control the peak of the stellar mass distribution. 
Because we have found that the mean temperature in molecular clouds and in high-mass star-forming cores may be controlled by
turbulent heating, the characteristic stellar mass may be affected by turbulent heating as well. For example, the larger gas temperature
predicted in high-mass star-forming cores may partly offset their large density, resulting in almost the same characteristic stellar mass,  
with respect to molecular clouds following Larson's relations. Finally, because the process of turbulent heating results in broad gas 
temperature distributions, it may be crucial in many molecular chemical reactions, besides those responsible for the formation of
CH$^+$ molecules.

\acknowledgements

This research was partially supported by a NASA ATP grant NNG056601G, and by
an NRAC allocation MCA098020S. We utilized computing resources provided by the San Diego Supercomputer 
Center, by the National Center for Supercomputing Applications and by NASA High End Computing Program.

\bibliographystyle{apj}

\begin{thebibliography}

\bibitem[Anselmet et al. 1984]{ans84}Anselmet, F., Gagne, Y., Hopfinger, E. J. \& Antonia, R. A.\ 1984, J. Fluid Mech. 140, 63.  
\bibitem [Arons \& Max 1975]{Aro75} Arons, J. \& Max, C. E. 1975, \apj, 196, 177L
\bibitem [Basu \& Murali 2001]{bas01}Basu, S. \& Murali, C. 2001, \apj, 551, 743 
\bibitem [Black 1987]{bla87}Black, J. H. 1987, in Interstellar processes, eds. Hollenbach, D. J. and Thronson, H. A., p731 
\bibitem [Boldyrev et al. 2002]{bol02} Boldyrev, S., Nordlund, A. \& Padoan, P. 2002, Phys. Rev. Lett., 89, 031102 
\bibitem [Bottorff \& Ferland 2002]{bot02}Bottorff, M. \& Ferland, G. 2002, \apj, 568, 581
\bibitem [Caselli et al. 1998]{cas98}Caselli, P., Walmsley, C. M., Terzieva, R., \& Herbst, E. 1998, \apj, 499, 234
\bibitem [Crane et al. 1995]{cra95}Crane, P., Lambert, D. L., \& Sheffer, Y. 1995, \apjs, 99, 107
\bibitem [Dalgarno \& McCray 1972]{dal72}Dalgarno, A. \& McCray, R. A. 1972, ARAA 10, 375
\bibitem [Dennis \& Chandran 2005]{den05}Dennis, T. J. \& Chandran, B. D. G. 2005, \apj, 622, 205
\bibitem [Draine \& Lee 1984]{dra84}Draine, B. T. \& Lee, H. M. 1984, \apj, 285, 89
\bibitem [Dubrulle 1994]{dub94} Dubrulle, B. 1994, Phys. Rev. Lett., 73,959
\bibitem [Evans 1999]{eva99}Evans, N. J. 1999, ARAA, 37, 311
\bibitem [Falgarone \& Puget 1995]{fal95}Falgarone, E. \& Puget, J.-L. 1995, \aap, 293, 840
\bibitem [Frisch 1995]{fri95} Frisch, U. 1995, Turbulence. (Cambridge University Press)
\bibitem [Goldreich \&  Kwan 1974]{gol74} Goldreich, P. \& Kwan, J. 1974, \apj, 189, 441
\bibitem [Goldsmith 2001]{gol01}Goldsmith, P. F. 2001, \apj, 557, 736
\bibitem [Goldsmith and Langer 1978]{gol78}Goldsmith, P. F. \& Langer, W. D.  1978, \apj, 222, 881
\bibitem [Gredel 1997]{gre97}Gredel, R. 1997, \aap, 320, 929
\bibitem [Gredel et al. 1993]{gre93}Gredel, R., van Dishoeck, E. F., \& Black, J. H. 1993, \aap, 269, 477
\bibitem [Habing]{hab68}Habing, H. J. 1968, Bull. Astron. Inst. Netherlands, 19, 421
\bibitem [Helie \& Troland 2003]{hel03}Heiles, C., \& Troland, T. H. 2003, \apj, 586, 1067
\bibitem [Hennebelle et al. 2007]{hen07}Hennebelle, P., Audit, E., \& Miville-Deschenes, M.-A. 2007, \aap, 465, 445
\bibitem [Heyer \& Brunt]{hey04}Heyer, M. H., Brunt, C. M. 2004, \apj, 615, L45 
\bibitem [Hollenbach \& McKee 1979]{hol79}Hollenbach, D. \& McKee, C. F. 1979, \apjs, 41, 555 
\bibitem [Kolmogorov 1962]{kol62}Kolmogorov, A. N. 1962, J. Fluid Mech. 13, 82
\bibitem [Kritsuk et al. 2007]{kri07}Kritsuk, A. G., Norman, M. L., Padoan, P., \& Wagner, R. 2007, \apj, 665, 416
\bibitem [Leurini et al. 2007]{leu07}Leurini, S., Schilke, P., Wyrowski, F., \& Menten, K. M. 2007, \aap, 466, 215
\bibitem [Liszt \& Lucas1996]{lis96}Liszt, H. \& Lucas, R. 1996, \aap, 314, 917
\bibitem [Lucas \& Liszt 1996]{luc96}Lucas, R. \& Liszt, H. 1996, \aap, 307, 237
\bibitem [Mac Low et al. 1998]{mac98}Mac Low, M.-M., Klessen, R. S., Burkert, A., \& Smith, M. D. 1998, phys. Rev. Lett., 80, 2754 
\bibitem [Mac Low 1999]{mac99}Mac Low, M.-M. 1999, \apj, 524, 169 
\bibitem [Matthaeus et al. 1999]{mat99}Matthaeus, W. H., Zank, G. P., Oughton, S., Mullan, D. J., \& Dmitruk, P. 1999, \apj, 523, L94
\bibitem [Minter \& Balser 1997]{min97}Minter, A. H. \& Balser, D. S. 1997, \apj, 484, 133
\bibitem [Mueller et al. 2002]{mue02}Mueller, K. E., Shirley, Y. L., Evans, N. J. \& Jacobson, H. R. 2002, \apjs, 143, 469
\bibitem [Muller \& Biskamp 2000]{mul00}Muller, W-C. \& Biskamp, D. 2000, Phys. Rev. Lett., 84, 475
\bibitem [Oboukhov 1962]{obo62}Oboukhov, A. M. 1962, J. Fluid Mech. 13, 77
\bibitem [Ossenkopf \& Henning 1994]{oss94}Ossenkopf, V. \& Henning, Th. 1994, \aap, 291, 943
\bibitem [Ossenkopf \& Mac Low 2002]{oss02}Ossenkopf, V. \& Mac Low, M.-M. 2002, \aap, 390, 307
\bibitem [Padoan \& Norlund 1999]{pad99} Padoan, P. \& Nordlund, A. 1999, \apj, 526, 279
\bibitem [Padoan et al. 2004]{pad04}Padoan, P., Jimenez, R., Nordlund, A., \& Boldyrev, S. 2004, Phys. Rev. Lett. 2004, 92, 191102
\bibitem [Padoan et al. 2006]{pad06} Padoan, P., Juvela, M., Kritsuk, A., \& Norman, M.~L.\ 2006, \apjl, 653, L125
\bibitem [Padoan et al. 2007]{pad07}Padoan, P., Nordlund, A., Kritsuk, A. G., Norman, M. L., Li, P. S. 2007, IAUS, 237, 283
\bibitem [Pan et al. 2008]{pan08}Pan, L., Wheeler, J. C., \& Scalo, J. 2008, \apj, accepted (astro-ph/08031689)
\bibitem [Pety \& Falgarone 2000]{pet00}Pety, J. \& Falgarone, E. 2000, \aap, 356, 279
\bibitem [Plume et al. 1997]{plu97}Plume, R., Jaffe, D. T., Evans, N. J., Martin-Pintado, J., \& Gomez-Gonzalez, J. 1997, \apj, 476, 730
\bibitem [She \& Leveque 1994]{she94}She, Z-S. \& Leveque, E. 1994, Phys. Rev. Lett, 72, 336
\bibitem [She \& Waymire 1995]{she95}She, Z-S. \& Waymire, E. C. 1995, Phys. Rev. Lett, 74, 262
\bibitem [Shirley et al. 2003]{shi03}Shirley, Y. L., Evans, N. J., Young, K. E., Knez, C., \& Jaffe, D. T. 2003, \apjs, 149, 375
\bibitem [Stone et al. 1998]{sto98}Stone, J. M., Ostriker, E. C., \& Gammie, C. F. 1998, \apj, 508, 99
\bibitem [van der Tak and van Dishoeck 2000]{van00} van der Tak, F. F. S. \& van Dishoeck, E. F. 2000, \aap, 358, 79L
\bibitem [Wolfire et al. 2003]{wol03}Wolfire, M. G., McKee, C. F., Hollenbach, D., \& Tielens, A. G. G. M. 2003, \apj, 587, 278
\bibitem [Yaglom 1966]{yag66}Yaglom, A. M. 1966, Dokl. Akad. Nauk SSSR, 166, 49  
\bibitem [Yeung et al. 2006]{yeu06}Yeung, P. K., Pope, S. B., Lamorgese, A. G. \&Donzis, D. A. 2006, Phys. Fluids. 18, 065103  
\end{thebibliography}

\end{document}